\documentclass[aps,prb,onecolumn,superscriptaddress,footinbib]{revtex4}
\usepackage{graphicx}
\usepackage{units}

\usepackage{eqnarray,amsmath}
\usepackage{graphicx}
\usepackage{alltt}
\usepackage[latin1]{inputenc}

\begin{document}
\title{Fourier Transform Analysis of STM Images of Multilayer Graphene Moiré Patterns}
\author{Frédéric Joucken}
\email[Corresponding author: ]{frederic.joucken@unamur.be}
\author{Fernande Frising}
\author{Robert Sporken}
\affiliation{Research Center in Physics of Matter and Radiation (PMR), University of Namur, 61 Rue de Bruxelles, 5000 Namur, Belgium}
\begin{abstract}
With the help of a simple model, we analyze Scanning Tunneling Microscopy images of simple and double moiré patterns resulting from misoriented bi- and tri-layers graphene stacks. It is found that the model reproduces surprisingly well non-trivial features observed in the Fast Fourier Transform of the images. We point out difficulties due to those features in interpreting the patterns seen on the FFT.
\end{abstract}
\maketitle
\section{Introduction}
Fast Fourier Transform (FFT) of Scanning Tunneling Microscopy (STM) images is a very common tool to gain complementary information about graphitic surfaces.\cite{xhie93,cee95,pong05bis,brihuega08,biedermann09,xue11,mallet12} Simple moiré patterns resulting from a misorientation between two graphene layers have already been studied extensively using FFT\cite{pong05bis,varchon08,wong10} but, although first studied a long time ago,\cite{xhie93} FFT studies of double moiré patterns on graphitic surfaces (due to the misalignment of three graphene layers) are still very scarce. 
\\
Here, we use a simple model for the electronic structure of graphene first intorduced by Hentschke {\textit{et al.}}\cite{hentschke92} to simulate STM images of bilayer rotated graphene displaying moiré patterns and discuss FFT non trivial signatures that are well reproduced by the model when we compute {\textit{constant current}} images. We then apply the model to the three-layer case and show that those non trivial features can hinder a proper interpretation of the FFTs. 
\section{Experiments and modeling}
Multilayer ($\sim$~5-10) graphene samples were obtained on SiC(000$\bar{1}$) by annealing the substrates in UHV at $\sim$~1320$^{\circ}$C for 12~min under a silicon flux of $\sim$~1ML/min.\cite{vanbommel75,forbeaux00} (The silicon flux is known to enlarge the graphene terraces by allowing higher growth temperature.\cite{tromp09}) This type of sample is known to contain grains that are not Bernal-stacked but twisted. This results in typical moiré patterns on STM images.\cite{varchon08,miller10} STM images have been obtained with a VP2 STM from Park Instrument at room temperature in UHV ($P\sim10^{-10}$~mbar) with electro-chemically etched tungsten tips. Unless stated otherwise, all experimental and simulated STM images presented below contain 512$^2$ points. 
\\
We use the model first introduced in ref\cite{hentschke92} where the local electronic density of a graphene layer is modeled, in a plane, by:
\[
\begin{split}
\phi(x,y) = 1&-\frac{2}{9}\left\{\cos\left[\frac{2\pi}{0.246}\left(x+y/\sqrt{3}\right)\right] \right.\\
  &\left.+ \cos\left[\frac{2\pi}{0.246}\left(x-y/\sqrt{3}\right)\right]\right.\\
  &\left.+\cos\left[\frac{4\pi}{0.246}\left(y/\sqrt{3}\right)\right]+\frac{3}{2}\right\}.
\end{split}
\]
This function represents well the local density of states of graphene near the Fermi level and is thus valid at small bias. To account for a rotation of $\theta$ between two layers, we compute $\phi(x,y)$ and $\phi(x',y')$ with
\begin{equation}
\left(
\begin{array}{c}
x'\cr
y'
\end{array}
\right)
=
\left(
\begin{array}{cc}
\cos\theta & -\sin\theta\cr
\sin\theta & \cos\theta
\end{array}
\right)
\left(
\begin{array}{c}
x\cr
y
\end{array}
\right).
\label{matrix}
\end{equation}
To evaluate the current, we simply include an exponential z-dependence and weight the contribution of each layer accordingly. This gives, for e.g. a three-layer sample:
\[
\begin{split}
I(x,&y,z)\propto\phi(x,y)e^{-z/\lambda}\\&+\phi(x',y')e^{-(z+d)/\lambda}+\phi(x'',y'')e^{-(z+2d)/\lambda}
\end{split}
\]
where $d$ is the interlayer distance (0.334~nm) and $\lambda$ the decay length. Although a realistic value of $\lambda$ is $\sim$0.05~nm,\cite{zhang08} we used a larger value ($\lambda=0.5$~nm) because we found that the convolution effects discussed below are best reproduced for this value.
\\
We recall here that the formula linking $D$, the periodicity of the moiré pattern and $\theta$, the angle between the two graphene layers producing this pattern is $D=a/[2\sin(\theta/2)]$ or $\theta=2\arcsin(2D/a)$.\cite{pong05bis}
\\
\textsc{Matlab} was used to compute the images and their FFTs while WSxM was used to treat both experimental and simulated images.\cite{horcas07} The $k$-convention in the FFTs is $k=1/l$ where $l$ is the length in the real-space image.
\section{Results and discussion}
\begin{figure}
\begin{center}
\includegraphics[width=0.8\columnwidth]{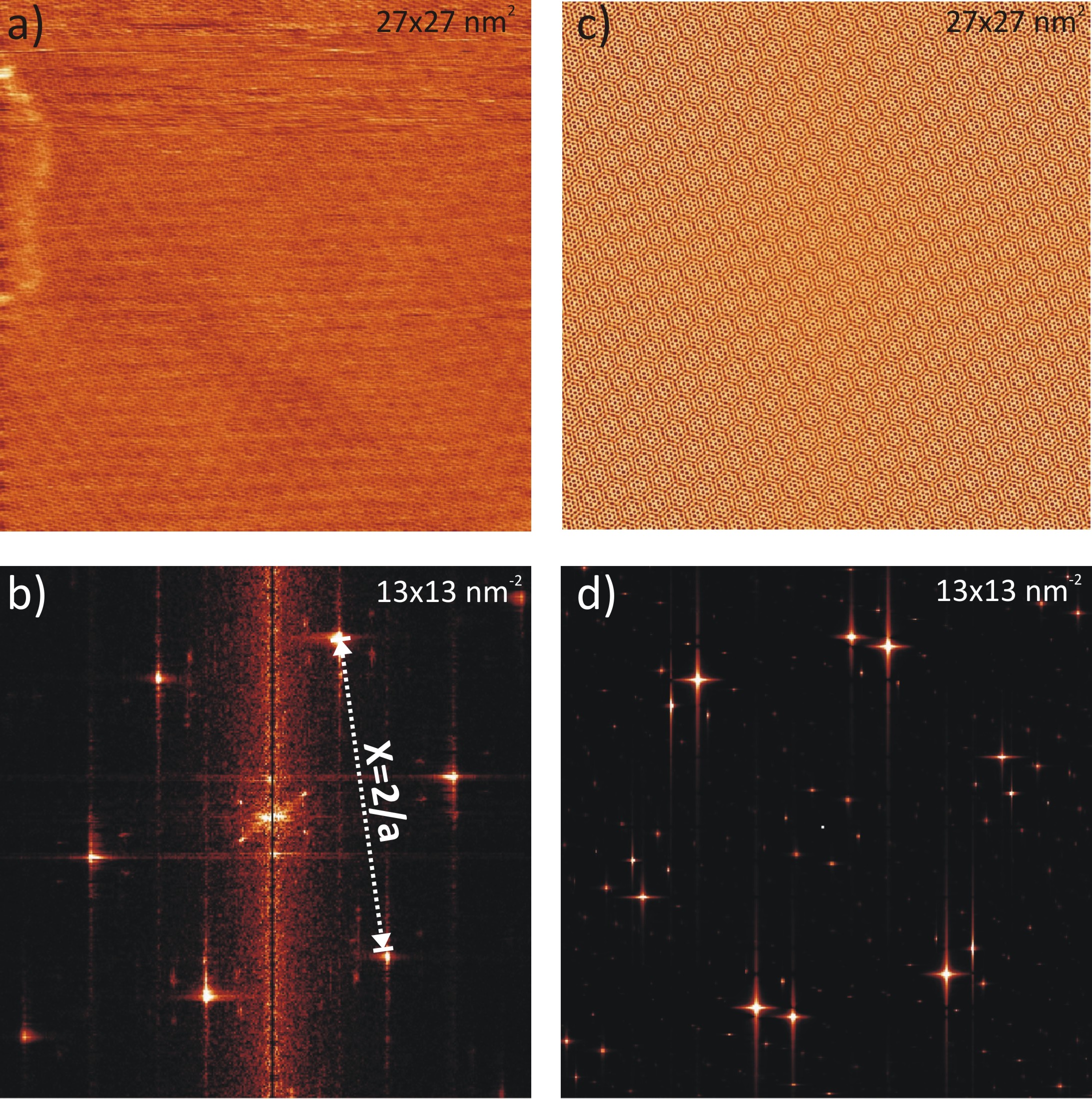}
\caption{a) Experimental constant-current STM image ($V_{sample}=-0.3$~V and $I=1$~nA) displaying a faint moiré. b) FFT of image a) c) Computed constant-current STM image of a twisted graphene bilayer with a misorientation angle $\theta=11.18^{^\circ}$ d) FFT of image c).}
\label{fig1}
\end{center}
\end{figure}

Fig.~\ref{fig1}a shows a constant-current experimental STM image ($V_{sample}=-0.3$~V and $I=1$~nA) atomically resolved and displaying a weakly contrasted moiré pattern. The FFT of this image is shown on Fig.~\ref{fig1}b. 
Besides the six points corresponding to the graphene lattice (note that to extract the lattice constance $a$ from the FFT, one should compute $(X/2)^{-1}$ where $X$ is the distance between two points shown on Fig.~\ref{fig1}b and not the distance between two opposite vertices of the hexagon) one can see an inner hexagon on the FFT whose dimensions match the periodicity of the moiré pattern on the real image: we find $D=1.26\pm0.03$~nm using the FFT and $D=1.27\pm0.03$~nm on the real-space image (the thermal drift due to the room temperature conditions limits significantly our precision). The presence of the moiré pattern in the real-space image does not imply the appearance of this hexagon in the FFT since the moiré could appear if the height variations of the tip would simply be the {\textit{addition}} of the height variations due to the two layers taken separately, in which case no signal having the periodicity matching the moiré's would appear in the FFT. So one could think that the presence of the inner hexagon in the FFT in Fig.~\ref{fig1}b is a sign of a topographic effect (the height of the atoms of the top layer is distributed by the moiré lattice). It is actually not obvious as this pattern is reproduced when we compute a {\textit{constant-current}} image of twisted bilayer graphene ($\theta=11.18^{^\circ}$), shown on Fig.~\ref{fig1}c together with its FFT on Fig.~\ref{fig1}d. (The constant-current image is obtained by calculating the tip's height producing a given setpoint current at each point of the image.)
\\
One also notices that the FFT of the computed image reproduces the replica of the inner hexagon around the six points of the graphene lattice seen on the FFT of the experimental image. This a sign of a convolution between the signal of the moiré and the signal of the graphene top layer {\textit{i.e.}} the presence of replica is due to a multiplication of the two signals in real-space.\cite{oranbrigham74} This is not surprising if one recalls that the moiré consists of AA-stacking (greater corrugation) succeeding to AB-stacking (smaller corrugation) with the moiré's periodicty {\textit{i.e.}} the amplitude of the corrugation varies with the moiré's periodicity.
\\
We now move on to the multiple moiré analysis. 
\begin{figure}
\begin{center}
\includegraphics[width=0.8\columnwidth]{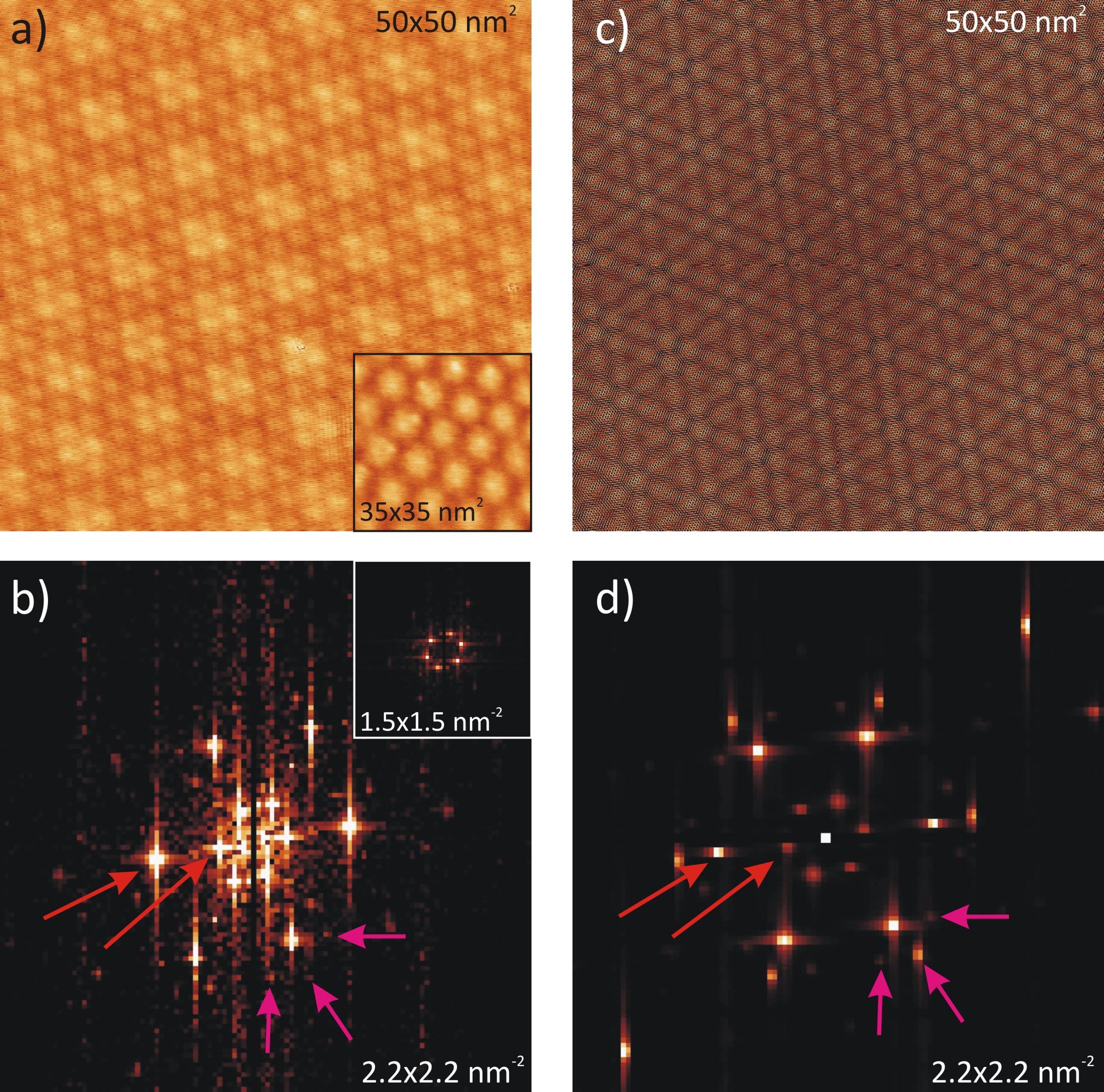}
\caption{a) Experimental STM image ($V_{sample}=-0.1$~V and $I=10$~nA) of a double moiré pattern (between three graphene layers) with $D_{12}=2.6$~nm ($\theta_{12}=5.5^{\circ}$) and $D_{23}=7.3$~nm ($\theta_{23}=1.9^{\circ}$). Inset: experimental STM image ($V_{sample}=+0.1$~V and $I=5$~nA) of a double moiré pattern with inverted misorientation angles ($\theta_{12}=1.8^{\circ}$ and $\theta_{23}=5.3^{\circ}$). b) Zoom in the FFT of a) on which replicas of the inner hexagon are seen around each vertex of the outer hexagon (one of them is pointed at by magenta arrows). Inset: zoom in the FFT of the image in the inset of a). c) Simulated image of a three layer stack with corresponding misorientation angles ($\theta_{12}=5.5^{\circ}$ and $\theta_{23}=1.9^{\circ}$). d) zoom in the FFT of c).}
\label{fig3}
\end{center}
\end{figure}
Fig.~\ref{fig3}a displays an STM image ($V_{sample}=-0.1$~V and $I=10$~nA) of a double moiré pattern resulting from imaging three misoriented graphene layers while Fig.~\ref{fig3}b shows the center of its FFT. As expected from the simple moiré case, the FFT shows two inner hexagons (pointed at by red arrows) that correspond to the two moiré patterns. We can deduce from those the periodicities of the moiré patterns: $D_{12}=2.6\pm0.3$~nm and $D_{23}=7.3\pm0.6$~nm {\textit{i.e.}} misorientation angles $\theta_{12}=5.5^{\circ}\pm 0.7^{\circ}$ and $\theta_{23}=1.9^{\circ}\pm 0.2^{\circ}$ ($D_{12}$ and $\theta_{12}$ and $D_{23}$ and $\theta_{23}$ refer to periodicities of the moiré patterns and angles between the first and the second layer and between the second and the third layer, respectively). We attribute the small periodicity to the moiré between the first and the second layer as we found other regions where double moirés with practically the same periods were present but with the misorientation angles $\theta_{12}$ and $\theta_{23}$ clearly inverted: an STM image ($V_{sample}=+0.1$~V and $I=5$~nA) of such a zone (for which we find $\theta_{12}=1.8^{\circ}\pm 0.2^{\circ}$ and $\theta_{23}=5.3^{\circ}\pm 0.3^{\circ}$) is shown in the inset of Fig.~\ref{fig3}a with a zoom in its corresponding FFT in the inset of Fig.~\ref{fig3}b. 
\\
Fig.~\ref{fig3}c displays a computed image of a trilayer stack with misorientation angles  $\theta_{12}=5.5^{\circ}$ and  $\theta_{12}=1.9^{\circ}$ while Fig.~\ref{fig3}d shows a zoom in the central part of its FFT.
\\
When looking at the FFT of the experimental image more closely, one can notice a similar phenomenon as in the bilayer case but now involving the moirés themselves: although not well-resolved, replicas of the smaller hexagon are found around the vertices of the greater hexagon. One of these replicas is pointed at by magenta arrows on Fig.~\ref{fig3}b. Interestingly, those replicas are also found in the FFT of the computed image (also pointed at by magenta arrows; the red arrows points at the two hexagons corresponding to the moirés). As in the previous case, they are also simply the result of the imaging mode in the computed case and are thus, in the real case, probably partly related to the imaging mode as well.
\\
We believe that those replicas can lead to misinterpretations of multiple moiré FFTs. 
\begin{figure}
\begin{center}
\includegraphics[width=0.8\columnwidth]{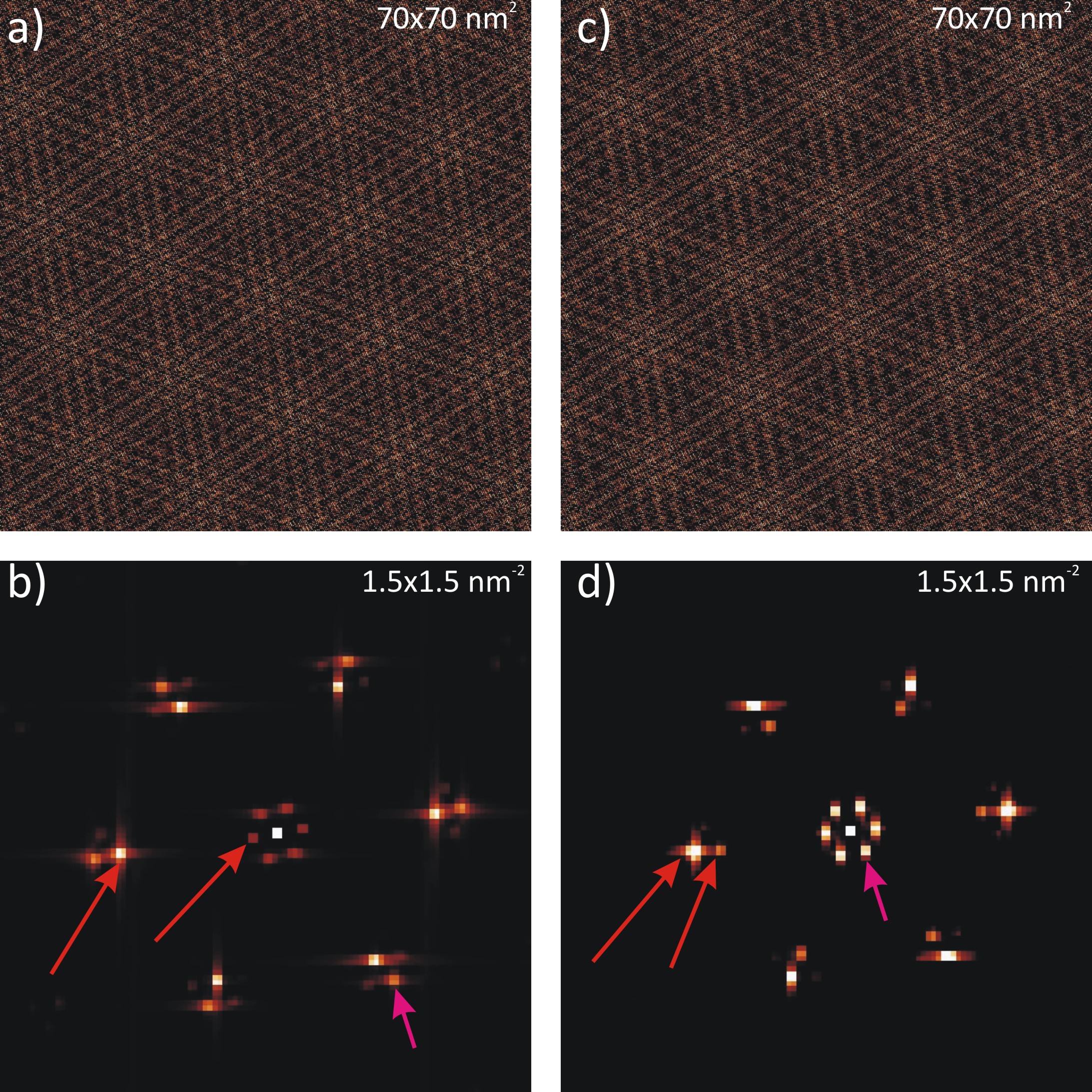}
\caption{a) Simulated STM image (1024$^2$ points) of trilayer graphene with misorientation angles $\theta_{12}=5.5^{\circ}$ and $\theta_{23}=0.9^{\circ}$. b) FFT of a) displaying the spots of the two moiré patterns pointed at with red arrows together with replicas of the long wavelength moiré near the spots of the short wavelength moiré (magenta arrow). c) Simulated STM image (1024$^2$ points) of trilayer graphene with misorientation angles $\theta_{12}=5.5^{\circ}$ and $\theta_{23}=-4.6^{\circ}$. d) FFT of c) displaying the spots of the two moiré patterns pointed at with red arrows; signs of interferences between the two moiré patterns are visible at the center (magenta arrow).}
\label{fig4}
\end{center}
\end{figure}
Indeed, Fig.~\ref{fig4}a shows a simulated STM image of a trilayer stack with misorientation angles $\theta_{12}=5.5^{\circ}$ and $\theta_{23}=0.9^{\circ}$ (a) together with a zoom in the central part of its FFT (b). On the FFT, the brightest sets of points are two hexagons that could be interpreted as corresponding to two moiré patterns whereas, if the inner one corresponds as expected to the moiré pattern between the first two layers, the outer one is actually a bright replica of the faint small hexagon near the center of the FFT that corresponds to the long wavelength moiré pattern. This situation could thus be misinterpreted as the presence of a double moiré but with wrong misorientation angles (in experiments, the long wavelength moiré spots located close to the center in the FFT can be lost in noise) or as a triple moiré.
\\
To illustrate the possible misinterpretation, a simulated image of a trilayer stack with misorientation angles $\theta_{12}=5.5^{\circ}$ and $\theta_{23}=-4.6^{\circ}$ ($\theta_{12}+\theta_{23}=0.9$) is presented on Fig.~\ref{fig4}c and a zoom on the central part of its FFT is shown on Fig.~\ref{fig4}d. The FFT displays the two sets of points corresponding to the two moiré patterns (pointed at by red arrows) but a central hexagon appears, with a corresponding wavelength of $\sim$~16~nm \textit{i.e.} the wavelength of a moiré pattern produced by a misorientation angle of $0.9^{\circ}$ (this long wavelength modulation is clearly visible on the image of Fig.~\ref{fig4}c; image which is practically indistinguishable from the case $\theta_{12}=5.5^{\circ}$ and $\theta_{23}=0.9^{\circ}$ of Fig.~\ref{fig4}a). As in the case of the simple moiré patterns (Fig.~\ref{fig1}), the presence of this central feature is not trivial and is the result of the imaging mode simulated here.
\\
A very similar situation has been reported recently on samples similar to ours and has been interpreted in terms of two misorientation angles very close to each other.\cite{miller10} The analysis presented above suggests that the interpretation of multiple moiré patterns FFTs might be less straightforward than thought.
\\
In conclusions, we reproduced subtle features in the FFTs of STM images of misoriented bi- and tri-layer graphene with the help of a model neglecting the possible interactions between the layers, indicating that those features cannot be straightforwardly interpreted as signs of non-purely electronic effects. We pointed out that FFTs of trilayer moiré patterns could easily be misinterpreted as having the wrong misorientation angles or as quadrilayer moiré patterns. To deepen the study, imaging similar samples in the constant-height mode and at low temperature would be of great interest.	
\section{Acknowledgements}
F. J. would like to thank Philippe Lambin for useful discussions and Etienne Gennart for technical support.
\def\bibfont{\footnotesize}

\end{document}